\documentclass[12pt]{article}
\usepackage{epstopdf}
\usepackage{epsfig}
\usepackage{graphicx}
\usepackage{a4}
\usepackage{amsmath}
\usepackage{latexsym}
\usepackage{cite}

\usepackage{color}
\usepackage{colordvi}

\DeclareGraphicsExtensions{.eps,.ps}

\usepackage{pslatex}
\usepackage[latin1]{inputenc}
\usepackage[T1]{fontenc}

\newcommand{\gsim}{\raisebox{-0.07cm}{$\:\:\stackrel{>}{{\scriptstyle \sim}}\:\: $} }

\newcommand{\msbar}{$\overline{\text{MS}}\, $}
\newcommand{\lp}{\left(}
\newcommand{\rp}{\right)}
\newcommand{\nn}{\nonumber}

\begin{document}

\begin{titlepage}
\noindent
DESY 13-092 \hfill May 2013\\
LPN 13-032 \\
SFB/CPP-13-36 \\
\vspace{1.3cm}

\begin{center}
  {\bf 
    \Large 
    Differential distributions for top-quark hadro-production \\[1ex]
    with a running mass
  }
  \vspace{1.5cm}

  {\large
    M. Dowling$^{\,a}$ 
    and
    S. Moch$^{\,a,b}$
  }\\
  \vspace{1.2cm}

  {\it 
    $^a$Deutsches Elektronensynchrotron DESY \\
    Platanenallee 6, D--15738 Zeuthen, Germany \\
    \vspace{0.2cm}
    $^b$ II. Institut f\"ur Theoretische Physik, Universit\"at Hamburg \\
    Luruper Chaussee 149, D--22761 Hamburg, Germany \\
  }
  \vspace{2.4cm}

\large
{\bf Abstract}
\vspace{-0.2cm}
\end{center}
We take a look at how the differential distributions for top-quark production
are affected by changing to the running mass scheme. 
Specifically we consider the transverse momentum, rapidity and pair-invariant
mass distributions at NLO for the top-quark mass in the \msbar\ scheme. 
It is found that, similar to the total cross section, the perturbative
expansion converges faster and the scale dependence improves using the mass in
the \msbar\ scheme as opposed to the on-shell scheme. 
We also update the analysis for the total cross section using the now
available full NNLO contribution. 
\vfill
\end{titlepage}

%
%
\newpage

The measurement of top-quark pair production cross sections at hadron colliders 
has entered the era of precision physics with the analysis of data available 
from the Large Hadron Collider (LHC) in the runs at center-of-mass energies $\sqrt{S}=7$ and $8$~TeV. 
Measurements of the total cross section for $t{\bar t}$-production from ATLAS and CMS 
reach by now an accuracy of typically better than ${\cal O}(10 \%)$, 
with the systematic and luminosity uncertainties already dominating over the
small statistical uncertainty, see, e.g.,~\cite{CMS:2012dya,ATLAS:2012jyc,CMS:2012gza}. 
First results of differential distributions for $t{\bar t}$-production from
the LHC are appearing as well~\cite{Chatrchyan:2012saa,Aad:2012hg}.
Thus, given the present experimental accuracy 
hadro-production of $t{\bar t}$-pairs is currently being established as a
Standard Model (SM) benchmark process.

This has motivated tremendous activity on the theory side to match the experimental precision by 
computing higher order corrections in Quantum Chromodynamics (QCD) and we briefly recapitulate 
the status for inclusive $t{\bar t}$-pair production, i.e., no additional jets or other tagged final states.
Predictions for the total cross section are complete to next-to-next-to-leading order 
(NNLO)~\cite{Baernreuther:2012ws,Czakon:2012zr,Czakon:2012pz,Czakon:2013goa}  
while differential distributions are known to next-to-leading order (NLO)~\cite{Beenakker:1988bq,Mangano:1991jk}, 
including top-quark decay~\cite{Melnikov:2009dn,Campbell:2012uf}, though.
Additional corrections beyond NLO based on threshold logarithms have been
obtained for distributions in the top-quark's transverse momentum and rapidity,
$p_T^t$ and $y^t$, as well as in the invariant mass $m^{t{\bar t}}$ of the top-quark pair~\cite{Kidonakis:2010dk,Ahrens:2009uz}.

Comparison of these theory predictions to experimental data can be 
used to determine non-perturbative parameters such as the strong coupling constant, 
the parton luminosity and the top-quark mass and to study their correlations.
Of these parameters, the top-quark mass is certainly the most interesting one with prominent 
implications for the electro-weak vacuum of the SM, see, e.g.,~\cite{Bezrukov:2012sa,Alekhin:2012py}. 
It is a particularly attractive feature of cross sections measurements that they offer the opportunity for
an unambiguous and theoretically well-defined determination of the top-quark mass 
in a particular renormalization scheme~\cite{Langenfeld:2009wd,Alioli:2013mxa}.

The conventional scheme choice for the quark mass renormalization is the pole mass, 
which has its short-comings~\cite{Bigi:1994em,Beneke:1994sw}, though, 
since it is based on the idea of quarks appearing as asymptotic states. 
It exhibits poor convergence of the perturbative series 
and due to the renormalon ambiguity it carries an intrinsic uncertainty 
of the order of $\Lambda_{\rm QCD}$.
As an alternative, one can consider top-quark hadro-production with a running mass, 
which has the advantages of improved convergence and scale stability of the perturbative expansion.
For $t{\bar t}$ hadro-production, these features have been demonstrated for the total cross section~\cite{Langenfeld:2009wd}.

In the present letter, we study the dependence of single differential distributions in $p_T^t$, $y^t$ and $m^{t{\bar t}}$ 
on the definition of the mass parameter. 
Specifically, we will compare the conventional pole mass $m_t^{\rm pole}$ with the scale dependent \msbar\ mass 
by means of the well-known relation in perturbation theory,
\begin{equation}
  \label{eq:onshtomsb}
  m_t^{\rm pole} \,=\, m(\mu_r)\left( 1 + \frac{\alpha_s}{\pi} d_1 + \left(\frac{\alpha_s}{\pi}\right)^2 d_2 
    + \dots \right),
\end{equation}
for the scheme change from $m_t^{\rm pole}$ to the running \msbar\ mass $m(\mu_r)$
taken at the renormalization scale $\mu_r$.
To NNLO the coefficients $d_1$ and $d_2$ are given by~\cite{Gray:1990yh} 
(see also Refs.~\cite{Chetyrkin:1999qi,Melnikov:2000qh}) 
\begin{eqnarray}
d_1 &=& \frac{4}{3} + \ell 
\\
d_2 &=& \frac{307}{32} + \frac{\pi^2}{3} + \frac{\pi^2}{9}\ln(2) - \frac{1}{6}\zeta_3 + \frac{509}{72}\ell + \frac{47}{24}\ell^2 
\\
& & - n_f \left( \frac{71}{144} + \frac{\pi^2}{18} + \frac{13}{36}\ell +
  \frac{1}{12}\ell^2 \right) 
\, ,
\nonumber
\end{eqnarray}
with $\ell = \ln\left(\frac{\mu_r^2}{m(\mu_r)^2}\right)$ and
assuming vanishing masses for all lighter quarks.

\begin{figure}[t!]
\centerline{
  \includegraphics[width=8.5cm]{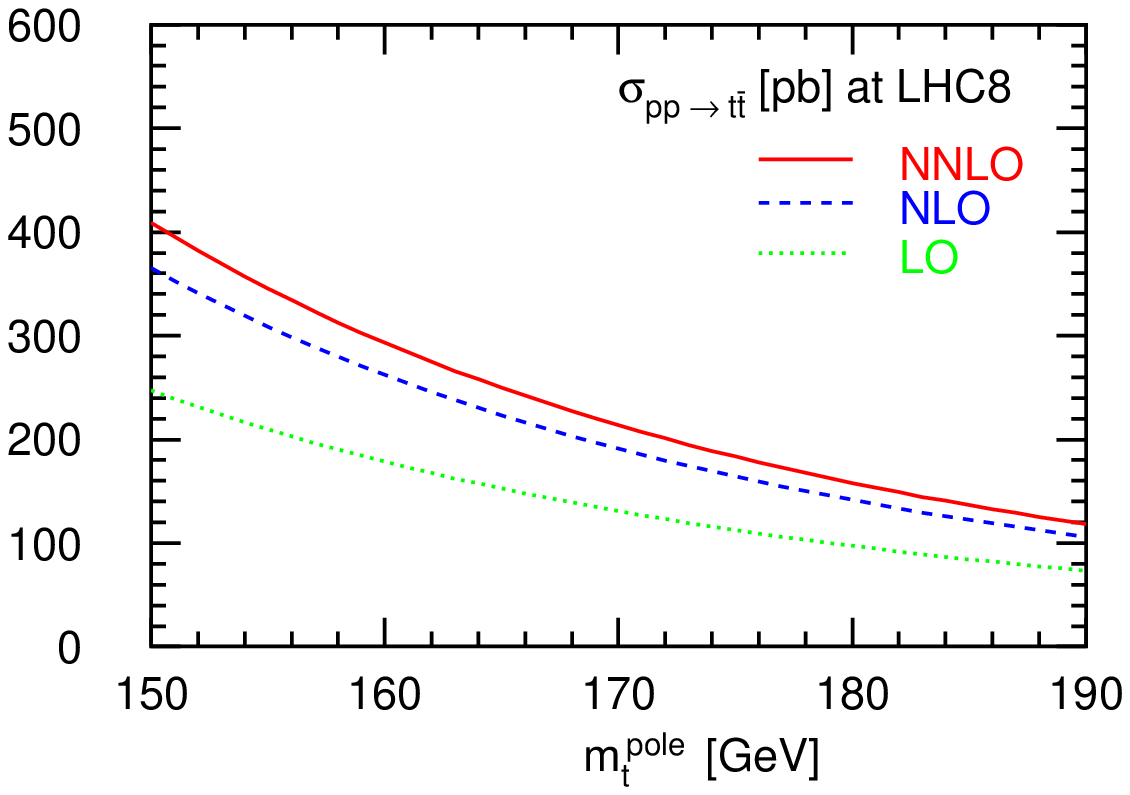}
  \includegraphics[width=8.5cm]{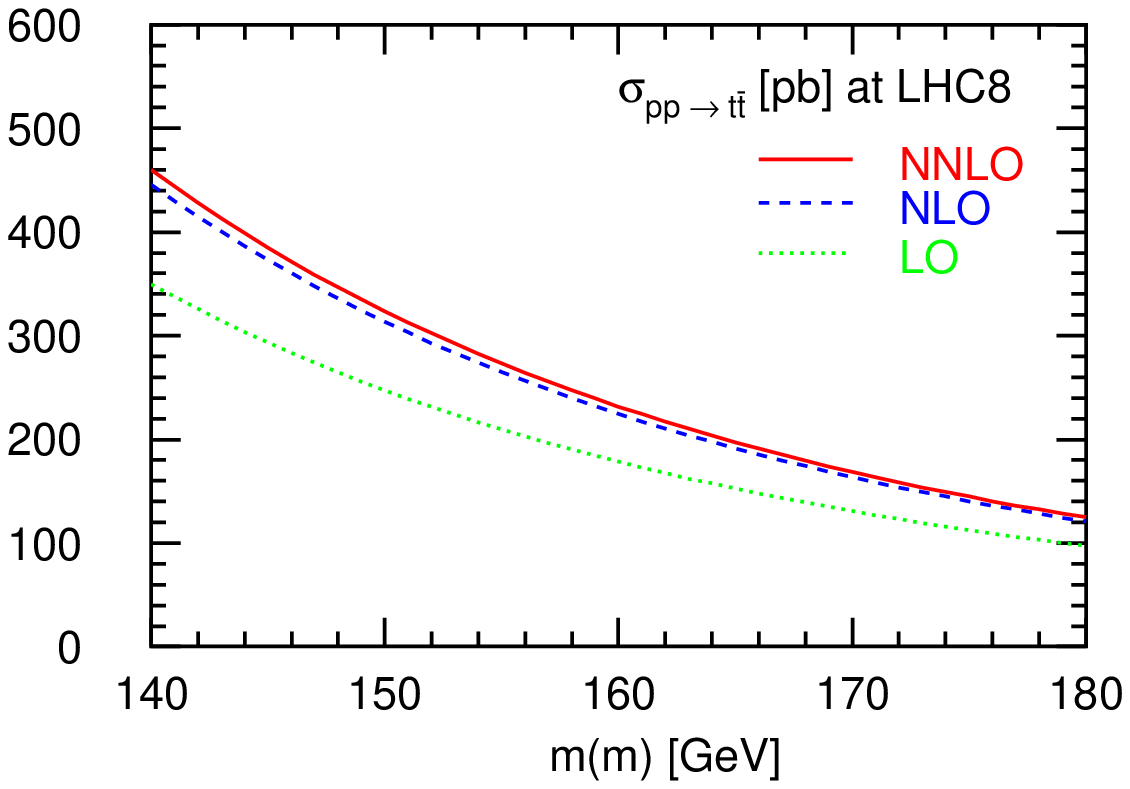}}
  \caption{\small
    \label{fig:sigmatot-mass}
    The  LO, NLO and NNLO QCD predictions for the 
    total cross section at LHC ($\sqrt{S} = 8$~TeV) 
    as a function of the top-quark mass  in the on-shell
    scheme $m_t^{\rm pole}$ at the scale $\mu = m_t^{\rm pole}$ (left) 
    and, respectively, in the \msbar\ scheme $m(m)$ at the scale $\mu = m(m)$ (right) 
    using the PDF set ABM11~\cite{Alekhin:2012ig} and $\mu = \mu_r = \mu_f$.
  }
\vspace*{10mm}
\centerline{
  \includegraphics[width=8.5cm]{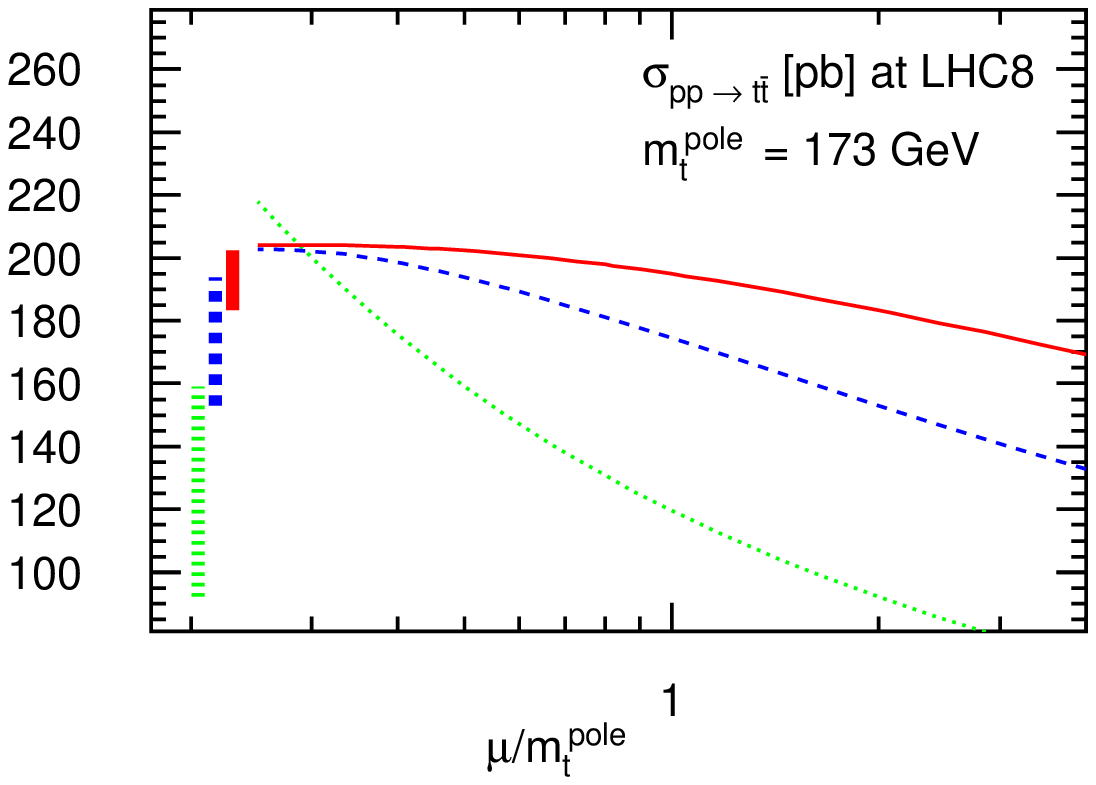}
  \includegraphics[width=8.5cm]{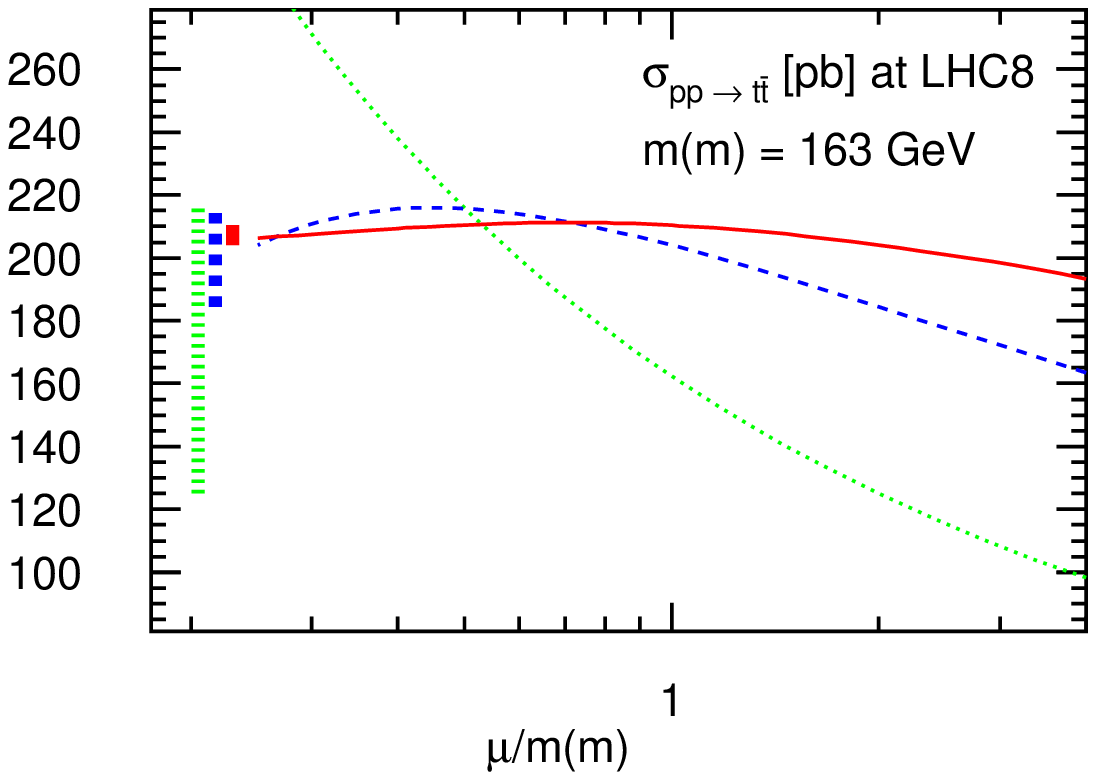}}
  \caption{\small
    \label{fig:sigmatot-mu}
    The scale dependence of the LO, NLO and NNLO QCD predictions for the 
    total cross section at LHC ($\sqrt{S} = 8$~TeV) 
    for the top-quark mass $m_t^{\rm pole}=173$~GeV in the on-shell scheme (left) 
    and for $m(m)=163$~GeV in the \msbar\ scheme (right) 
    with the choice $\mu = \mu_r = \mu_f$ 
    using the PDF set ABM11~\cite{Alekhin:2012ig}. 
    The vertical bars indicate the size of the scale variation in the standard
    range $\mu/m_t^{\rm pole} \in [1/2, 2]$ and $\mu/m(m) \in [1/2, 2]$, respectively.
}
\end{figure}

\bigskip

Let us briefly illustrate the advantages of the running \msbar\ mass $m(\mu_r)$ for the total $t{\bar t}$ cross section.
The recently completed exact NNLO QCD result~\cite{Baernreuther:2012ws,Czakon:2012zr,Czakon:2012pz,Czakon:2013goa}
turned out to be very close, i.e., within ${\cal O}(1 - 2\%)$, 
to previous approximations based on the combined threshold and high-energy asymptotics~\cite{Moch:2012mk} 
and has been presented as a function of the pole mass $m_t^{\rm pole}$.
The necessary scheme transformation 
from $m_t^{\rm pole}$ to $m(\mu_r)$, i.e., the application of eq.~(\ref{eq:onshtomsb}), 
has been discussed in~\cite{Langenfeld:2009wd} and 
is implemented in the program {\tt Hathor} (version 1.5)~\cite{Aliev:2010zk},
a tool for the calculation of the total $t{\bar t}$ cross section in hadronic collisions.

The much improved apparent convergence of the perturbative expansion 
with the running mass as well as the scale stability 
are illustrated in Figs.~\ref{fig:sigmatot-mass} and \ref{fig:sigmatot-mu} 
where we compare theory predictions for the total $t{\bar t}$ cross section 
as a function of the pole and the \msbar\ mass, respectively.
Fig.~\ref{fig:sigmatot-mass} displays the increase in the cross section values 
from LO to NNLO, where we have taken the parton distribution functions (PDFs) 
to be order independent.
For an on-shell mass $m_t^{\rm pole}=173$~GeV, for instance, 
the relative increase is $\sigma_{\rm NLO}/\sigma_{\rm LO} = 1.46$ and $\sigma_{\rm NNLO}/\sigma_{\rm NLO} = 1.12$
at the scale $\mu_r = \mu_f = m_t^{\rm pole}$.
This is to be compared with a much reduced increase of only 
$\sigma_{\rm NLO}/\sigma_{\rm LO} = 1.26$ and $\sigma_{\rm NNLO}/\sigma_{\rm NLO} = 1.03$ 
for $m(m)=163$~GeV in the \msbar\ scheme at the scale $\mu_r = \mu_f = m(m)$.
These findings can be understood by noting that the scheme transformation of eq.~(\ref{eq:onshtomsb}) applied to the 
total $t{\bar t}$ cross section effectively shifts all parton-level corrections 
to the threshold region thereby improving the apparent convergence of the 
perturbation series, see, e.g., ~\cite{Moch:2010rh}.

Fig.~\ref{fig:sigmatot-mu} shows the scale stability for the LHC predictions 
confirming earlier findings for the Tevatron, cf.~\cite{Langenfeld:2009wd}.
The scale variation for the cross section in the on-shell 
scheme in the standard range $\mu/m_t^{\rm pole} \in [1/2, 2]$
amounts to $\Delta \sigma_{\rm NNLO} = ^{+3.8\%}_{-6.0\%}$, 
whereas for the running mass we only find $\Delta \sigma_{\rm NNLO} = ^{+0.1\%}_{-3.0\%}$ 
for the range $\mu/m(m) \in [1/2, 2]$.
Interestingly, for an on-shell mass the point of minimal sensitivity where 
$\sigma_{\rm LO} \simeq \sigma_{\rm NLO} \simeq \sigma_{\rm NNLO}$ is located
at fairly low scales, $\mu \simeq m_t^{\rm pole}/4 \simeq 45$~GeV, whereas 
for a running mass it resides at the scale $\mu = {\cal O}(m(m))$, i.e., 
it coincides with the natural hard scale of the process.
These results imply, that experimental determinations of the running mass from
the measured cross section are feasible with very good accuracy and a small 
residual theoretical uncertainty.
For Tevatron data such analyses have already been performed in the past~\cite{Abazov:2011pta,Alekhin:2012py}.

For completeness, we include here values for the full NNLO cross sections at
the Tevatron ($\sqrt{S}=1.96$~TeV) and at the LHC for various energies of interest.
\begin{table}[th!]
\renewcommand{\arraystretch}{1.3}
\begin{center}
{\small
\hspace*{-5mm}
\begin{tabular}{|l|l|l|l|l|l|}
\hline
&\multicolumn{1}{|c|}{TEV $\sqrt{S} = 1.96\, {\rm TeV}$ } 
&\multicolumn{1}{|c|}{LHC $\sqrt{S} = 7\, {\rm TeV}$ }
&\multicolumn{1}{|c|}{LHC $\sqrt{S} = 8\, {\rm TeV}$ }
&\multicolumn{1}{|c|}{LHC $\sqrt{S} = 14\, {\rm TeV}$ }
\\     
\hline
ABM11 &
$6.82~^{+0.21}_{-0.29}~^{+0.16}_{-0.16}$&
$133.0~^{+5.2}_{-8.2}~^{+6.5}_{-6.5}$&
$194.9~^{+7.4}_{-11.7}~^{+8.8}_{-8.8}$&
$821.0~^{+27.0}_{-43.7}~^{+25.7}_{-25.7}$
\\
CT10 & 
$7.30~^{+0.28}_{-0.39}~^{+0.45}_{-0.33}$&
$168.9~^{+6.9}_{-10.9}~^{+13.5}_{-10.9}$&
$241.6~^{+9.5}_{-15.1}~^{+16.9}_{-13.8}$&
$939.3~^{+32.4}_{-51.7}~^{+37.5}_{-33.3}$
\\
\hline
\end{tabular}
}
\caption{\small 
The total cross section for top-quark pair-production at NNLO using a 
pole mass $m_t^{\rm pole} = 173~{\rm GeV}$ and the PDF set
ABM11~\cite{Alekhin:2012ig} and CT10~\cite{Gao:2013xoa} and with 
the errors shown as $\sigma + \Delta \sigma_{\rm scale} + \Delta \sigma_{\rm PDF}$.
The scale uncertainty $ \Delta \sigma_{\rm scale}$ is based on maximal and minimal 
shifts for the choices $\mu=m_t^{\rm pole}/2$ and $\mu = 2m_t^{\rm pole}$ and 
$\Delta \sigma_{\rm PDF}$ is the 1$\sigma$ combined PDF+$\alpha_s$ error.
All rates are in pb. 
}
\label{tab:ttbar-pole}
\end{center}
\end{table}

\begin{table}[th!]
\renewcommand{\arraystretch}{1.3}
\begin{center}
{\small
\hspace*{-5mm}
\begin{tabular}{|l|l|l|l|l|l|}
\hline
&\multicolumn{1}{|c|}{TEV $\sqrt{S} = 1.96\, {\rm TeV}$ } 
&\multicolumn{1}{|c|}{LHC $\sqrt{S} = 7\, {\rm TeV}$ }
&\multicolumn{1}{|c|}{LHC $\sqrt{S} = 8\, {\rm TeV}$ }
&\multicolumn{1}{|c|}{LHC $\sqrt{S} = 14\, {\rm TeV}$ }
\\     
\hline
ABM11 &
$7.22~^{+0.10}_{-0.10}~^{+0.16}_{-0.16}$&
$143.8~^{+0.2}_{-4.3}~^{+6.4}_{-6.4}$&
$210.4~^{+0.1}_{-6.3}~^{+8.6}_{-8.6}$&
$880.0~^{+0.0}_{-24.0} ~^{+24.6}_{-24.6}$
\\
CT10 & 
$7.70~^{+0.10}_{-0.15}~^{+0.47}_{-0.35}$&
$180.7~^{+0.0}_{-5.8}~^{+13.7}_{-11.1}$&
$258.0~^{+0.0}_{-8.1}~^{+17.2}_{-14.1}$&
$997.9~^{+0.0}_{-28.3} ~^{+38.1}_{-33.9}$
\\
\hline
\end{tabular}
}
\caption{\small 
Same as Tab.~\ref{tab:ttbar-pole} for a running mass 
$m(m) = 163~{\rm GeV}$ in the \msbar\ scheme.
}
\label{tab:ttbar-msbar}
\end{center}
\end{table}

\bigskip

Next we discuss the single-differential distributions in the top-quark's transverse momentum $p_T^t$
and rapidity $y^t$ and in the invariant mass $m^{t{\bar t}}$ of the $t{\bar t}$-pair, 
which are all known to NLO in QCD~\cite{Beenakker:1988bq,Mangano:1991jk} 
in the conventional pole mass scheme.
As we are interested in the differential cross sections with the mass in the \msbar\ scheme,
we briefly recall the kinematics of heavy-quark hadro-production,
\begin{eqnarray}
  \label{eq:3}
  h_1(P_1) +\, h_2(P_2) &\longrightarrow& {\rm{Q}}(p_1) +\,
  X[{\Bar{\rm{Q}}}](p_X)\, , 
\end{eqnarray}
where $h_1$ and $h_2$ are hadrons, $X[\overline{\rm{Q}}]$ denotes any
allowed hadronic final state containing at least the heavy anti-quark,
and ${\rm{Q}}(p_1)$ is the identified heavy-quark with mass $m$.  The
hadronic invariants in this reaction are
\begin{equation}
  \label{eq:5}
S = (P_1+P_2)^2 \quad,\quad T_1 = (P_2-p_1)^2-m^2 \quad,\quad U_1 = (P_1-p_1)^2-m^2
\, . 
\end{equation}

The double differential cross section for eq.~(\ref{eq:5}) in terms of the 
hard parton cross section $\sigma_{ij}$ and PDFs $f_i$ at the factorization scale $\mu^2$ reads 
\begin{equation}
  \label{eq:dbldff}
S^2\frac{d^2\sigma(S,T_1,U_1)}{dT_1 dU_1} = \int_{x_1^-}^1 \frac{dx_1}{x_1} \int_{x_2^-}^1 \frac{dx_2}{x_2} f_i(x_1,\mu^2) f_j(x_2,\mu^2) s^2 
\frac{d^2\sigma_{ij}(s,t_1,u_1,\mu^2)}{dt_1 du_1} 
\, ,
\end{equation}
and the partonic invariants are related to their hadronic counterparts through
\begin{equation}
t_1 = x_1 T_1 \, , \qquad\qquad u_1 = x_2U_1\, , \qquad \qquad s = x_1 x_2 S
\, ,
\end{equation}
with the limits on $x_1$ and $x_2$,
\begin{equation}
x_1^- = -\frac{U_1}{S+T_1} \leq x_1 \leq 1\, , \qquad\qquad x_2^- = \frac{x_1 T_1}{x_1S + U_1} \leq x_2 \leq 1
\, .
\end{equation}
In order to write the differential cross section in terms of $p_T^t$, $y^t$ and $m^{t{\bar t}}$, we will also need their definitions in terms of the hadronic invariants.
For the case of $p_T^t$ and $y^t$, the relations are
\begin{equation}
y^t = \frac{1}{2}\ln\left(\frac{T_1}{U_1}\right)\, , \qquad\qquad (p_T^t)^2=\frac{T_1U_1}{S}-m^2
\, ,
\end{equation}
whereas for $m^{t{\bar t}}$, pair-invariant mass kinematics is used, in 
which case the requirements on the integrals are
\begin{equation}
x_1^- = \frac{\left(m^{t{\bar t}}\right)^2}{S} \qquad \textrm{and} \qquad x_2^- = \frac{\left(m^{t{\bar t}}\right)^2}{x_1 S}
\, .
\end{equation} 
In these kinematics, the relevant partonic invariants for writing the differential cross section in terms of $m^{t{\bar t}}$ are,
\begin{eqnarray}
\label{eq:pimdefs}
t_1 = -\frac{\left(m^{t{\bar t}}\right)^2}{2}\left( 1-\beta_t \cos\theta \right)
\, , \qquad\qquad
u_1 = -\frac{\left(m^{t{\bar t}}\right)^2}{2}\left( 1+\beta_t \cos\theta \right) 
\, ,
\end{eqnarray}
with $\beta_t = \sqrt{1-4m^2/\left(m^{t{\bar t}}\right)^2}$ and $\theta$ the scattering angle of the top quark.
Full discussions of the kinematics to NLO for one-particle inclusive and pair-invariant mass kinematics are available in \cite{Beenakker:1990maa,Mangano:1991jk} respectively.

In order to convert to cross section predictions with the mass in the \msbar\ scheme,
we start from the on-shell description:
\begin{equation}
\frac{d\sigma(m_t^{\rm pole})}{dX} \,=\, 
\lp\frac{\alpha_s}{\pi}\rp^2 \frac{d\sigma^{(0)}(m_t^{\rm pole})}{dX} 
+ \lp\frac{\alpha_s}{\pi}\rp^3 \frac{d\sigma^{(1)}(m_t^{\rm pole})}{dX} + {\cal O}(\alpha_s^2)
\, ,
\end{equation}
where $X$ denotes any of the variables $p_T^t$, $y^t$ and so on.
If we now replace $m_t^{\rm pole}$ with $m(\mu_r)$ using eq.~(\ref{eq:onshtomsb}), 
we can expand in $\alpha_s$ and obtain a description of the differential cross section in the \msbar\ scheme.
\begin{eqnarray}
\label{eq:crosssecmsb}
\frac{d\sigma(m(\mu_r))}{dX} &=& 
\lp\frac{\alpha_s}{\pi}\rp^2\frac{d\sigma^{(0)}(m(\mu_r))}{dX} \\
& &  
+ \lp\frac{\alpha_s}{\pi}\rp^3 \left\{ 
  \frac{d\sigma^{(1)}(m(\mu_r))}{dX} 
  + d_1 m(\mu_r)
  \frac{d}{dm_t}\lp\frac{d\sigma^{(0)}(m_t)}{dX}\rp\biggr|_{m_t=m(\mu_r)}
\right\}
+ {\cal O}(\alpha_s^2)
\, .
\nn 
\end{eqnarray}
The only extra part required is the mass derivative of the Born contribution.
This has been computed semi-analytically for the $p_T^t$, $y^t$, and $m^{t{\bar t}}$ distributions.
To see why we also need some numerical derivatives in this calculation, consider eq.~(\ref{eq:dbldff}) 
for the Born contribution to the double differential cross section as a starting point:
\begin{equation}
\int_{x_1^-}^1 \frac{dx_1}{x_1} \int_{x_2^-}^1 \frac{dx_2}{x_2} f_i(x_1,\mu^2) f_j(x_2,\mu^2) s^2 
\frac{d^2\sigma_{ij}^{(0)}}{dt_1 du_1} \delta(s+t_1+u_1)
\, ,
\end{equation}
where the delta function imposes Born kinematics 
and can be used to carry out the integral over $x_2$ through its relation to
$s, t_1$ and $u_1$.
Re-writing the cross section in terms of $p_T^t$ and $y^t$ provides us with the form 
of the integrand that will need to be evaluated,
\begin{equation}
  \int_{x_1^-}^1 dx_1 {\cal L}(x_1,x_2,\mu^2) 
  \frac{x_1 x_2 S}{x_1 S + U_1} 
  \left.\frac{d^2\sigma(s,t_1,u_1)}{dy^t dp_T^{t2}}\right|_{x_2=-\frac{x_1 T_1}{x_1 S+U_1}}
  \, ,
\label{eq:integrand}
\end{equation}
where ${\cal L}(x_1,x_2,\mu^2) = f_1(x_1,\mu^2)f_2(x_2,\mu^2)/x_1x_2$ is the
differential parton luminosity.

The most important aspect to note is that both $x_2$ and $x_1^-$ 
depend on the top-quark mass through their relations to the Mandelstam variables.
This means that the mass derivative of the PDFs needs to be done numerically using
\begin{eqnarray}
\frac{d}{dm} {\cal L}(x_1,x_2,\mu^2)  & = & \frac{dx_1}{dm}\frac{{\cal L}(x_1+\delta,x_2,\mu^2) - {\cal L}(x_1-\delta,x_2,\mu^2)}{2\delta} \nonumber \\
& & + \frac{dx_2}{dm}\frac{{\cal L}(x_1,x_2+\delta,\mu^2) - {\cal L}(x_1,x_2-\delta,\mu^2)}{2\delta}
\, .
\end{eqnarray}
This form of the derivative is found to converge well.
Aside from this, all other derivatives are known analytically.
When compared with a fully numerical calculation of the derivative term, 
it is found that the two methods agree to less than 1\%. 
In the case of $m^{t{\bar t}}$, the integration limits and variables 
do not depend on the top-quark mass ($m$) so all derivatives are computed analytically.

\bigskip

Using the relations presented here, 
we have computed the differential cross sections for $t{\bar t}$-production in terms of $p_T^t$, $y^t$ and $m^{t{\bar t}}$.
We have used the program {\tt MCFM}~\cite{MCFM} for the NLO corrections~\cite{Campbell:2010ff,Campbell:2012uf} 
in the conventional pole mass $m_t^{\rm pole}$ scheme 
and a custom routine for the Born and mass derivative terms.
The calculations were carried out using the ABM11~\cite{Alekhin:2012ig} and CT10~\cite{Gao:2013xoa} PDFs at NLO.
As a check, each curve was integrated to obtain a result for the full cross section.
In all cases, the value agreed within less than 1\% of the cross section computed 
using {\tt Hathor}.
As well, the mass derivatives were checked by computing the differential cross sections 
at values of the top mass ranging between 150~GeV and 180~GeV.
A curve was fit to each point in the relevant spectrum to obtain the derivative at the given \msbar\  mass.
Again, these values agreed within less than 1\% of the (semi-)analytic derivatives used.

\begin{figure}[t!]
\includegraphics[width=8.5cm]{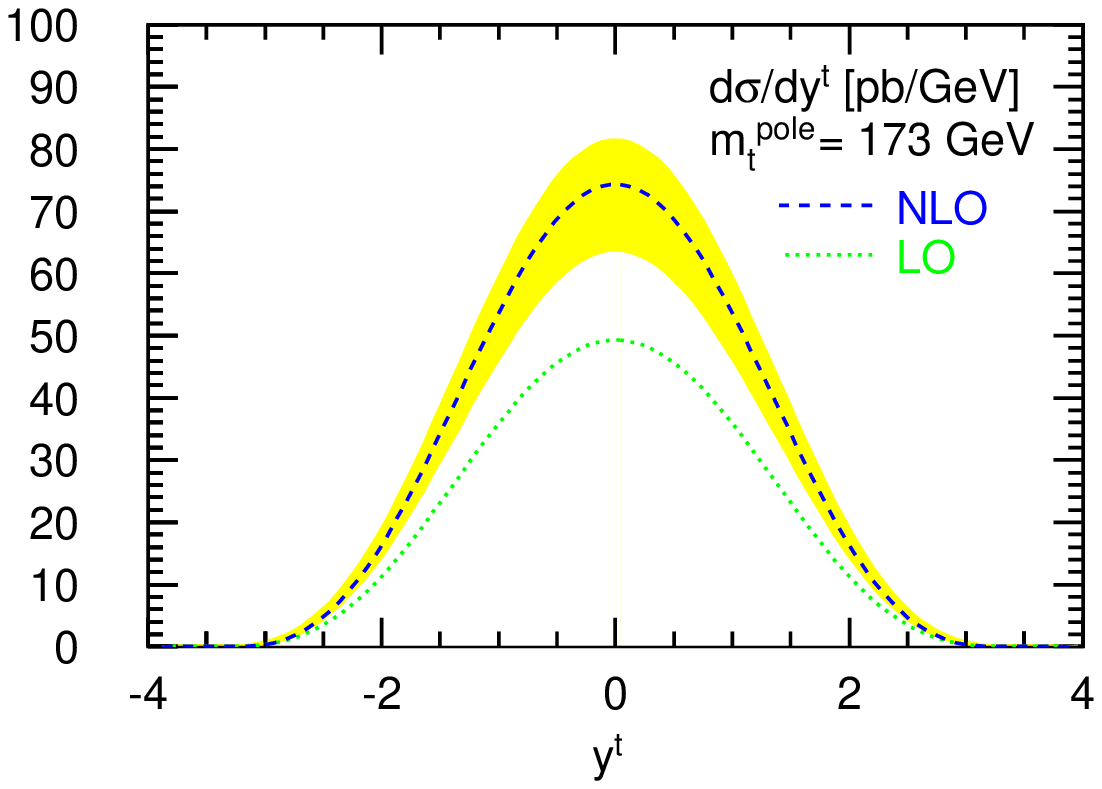}
\includegraphics[width=8.5cm]{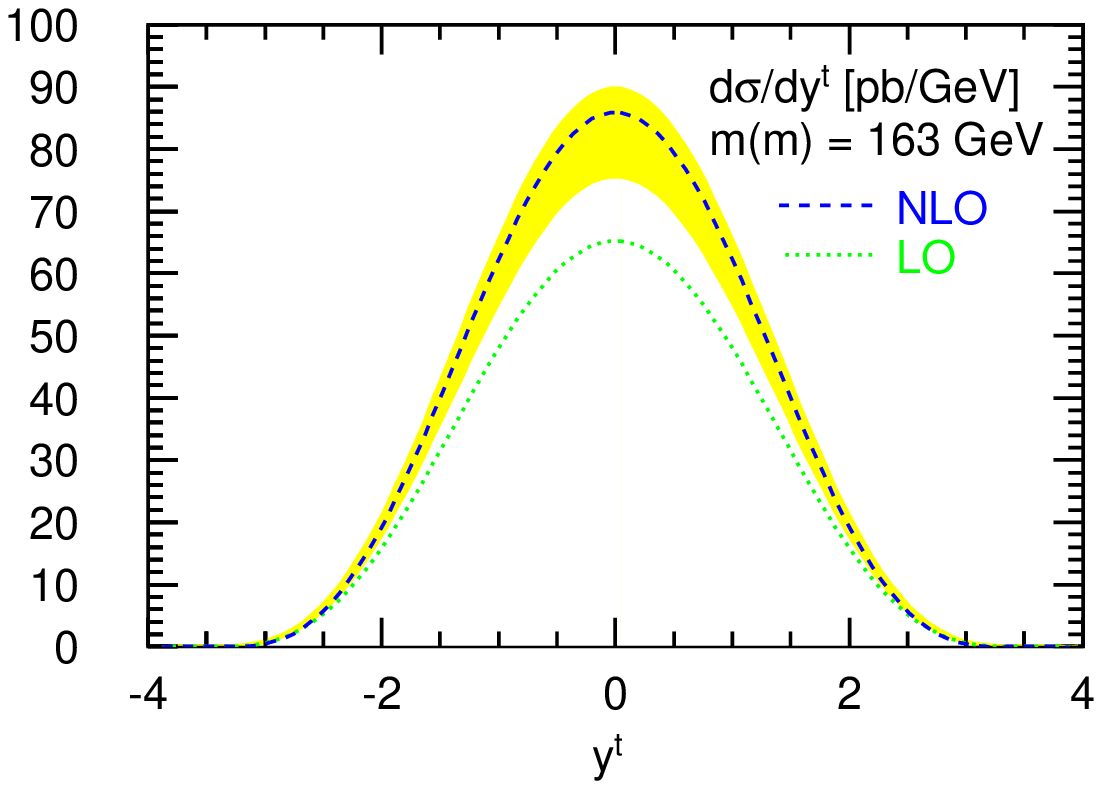}
\caption{\small \label{fig:yt} 
The differential cross section with respect to the rapidity $y^t$
of the top quark in the pole (left) and the \msbar\  (right) mass scheme at the LHC
with $\sqrt{S}=8$~TeV.
The dotted (green) curves are the LO contributions while the dashed (blue) 
curves include NLO corrections and are obtained using the PDF set CT10~\cite{Gao:2013xoa}.
The scale dependence in the range $\mu/m_t^{\rm pole}$ 
or $\mu/m(m) \in [1/2,2]$ is shown as a band around the NLO curve.
}
\end{figure}

In Fig.~\ref{fig:yt} the rapidity distributions are shown for the \msbar\  and pole mass schemes.
It is clear from these that at NLO, the convergence of the perturbative series as well as the scale dependence improves.
In the pole-mass scheme, a relative increase for the cross section
ratios $\sigma_{\rm NLO}/\sigma_{\rm LO} = 1.50$ is seen, while in the \msbar\  scheme we have $\sigma_{\rm NLO}/\sigma_{\rm LO} = 1.31$ at $y^t= 0$.
The scale variation in the on-shell scheme is 
$\Delta \sigma_{\rm NLO} = {}^{+9.5\%}_{-14\%}$ while in the \msbar\  scheme, we have $\Delta \sigma_{\rm NLO} = {}^{+4.5\%}_{-12\%}$ again at $y^t = 0$.

\begin{figure}[t!]
\includegraphics[width=8.5cm]{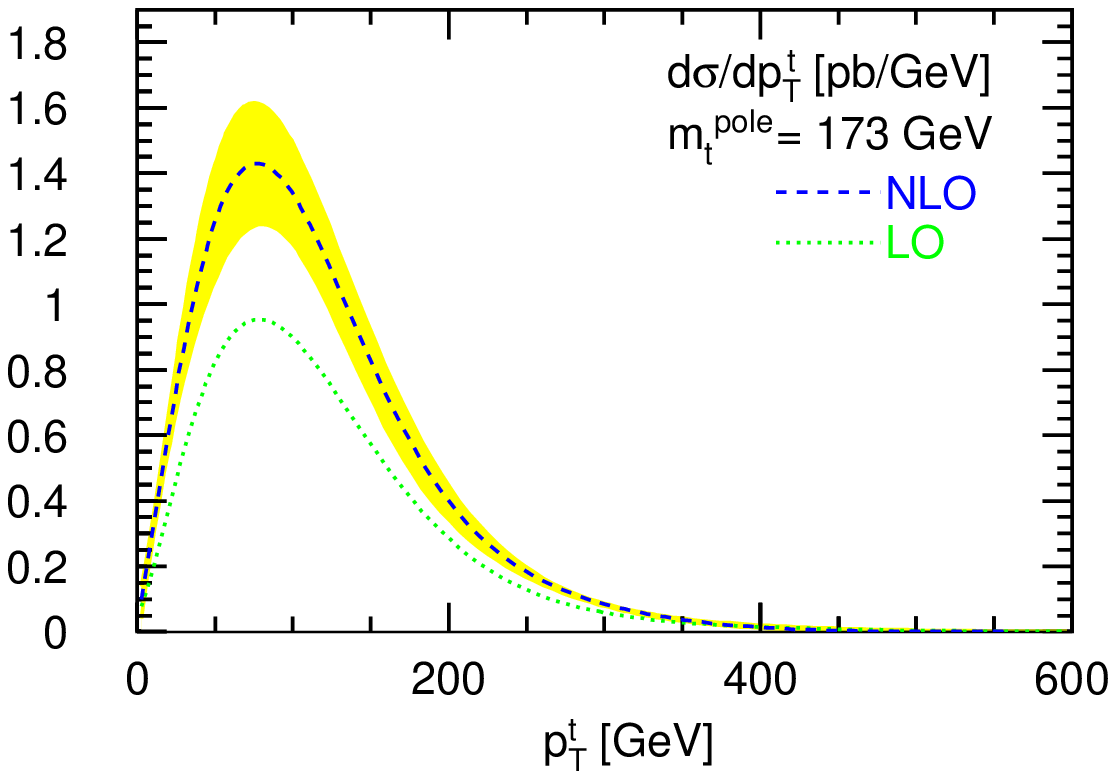}
\includegraphics[width=8.5cm]{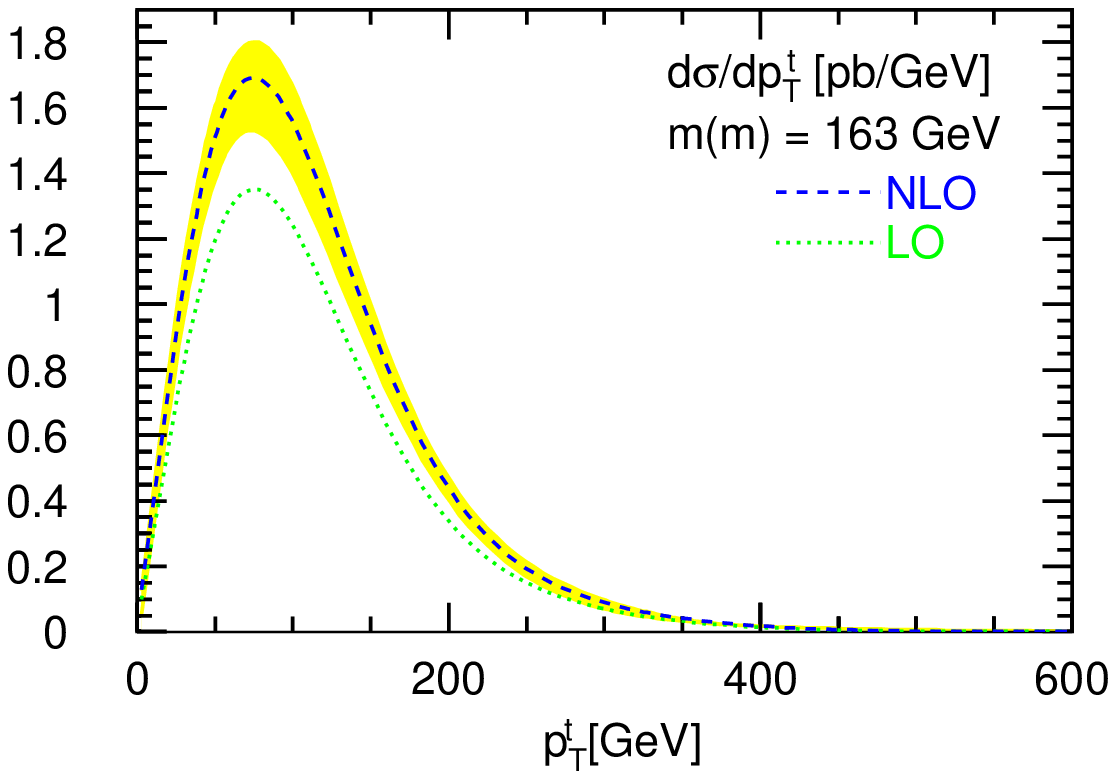}
\caption{\small \label{fig:pt}
Same as Fig.~\ref{fig:yt} for the differential cross section with
respect to the transverse momentum 
$p_T^t$ of the top quark.
}
\end{figure}

Fig.~\ref{fig:pt} shows the transverse momentum distributions.
Again we see an improvement when moving from the pole mass scheme to the \msbar\  scheme.
In this case the improvement in the NLO contribution is a bit better with
$\sigma_{\rm NLO}/\sigma_{\rm LO} = 1.50$ 
for the pole mass scheme and $\sigma_{\rm NLO}/\sigma_{\rm LO} = 1.25$ in the \msbar\  scheme.
The scale variation goes from $\Delta \sigma_{\rm NLO} = {}^{+13\%}_{-13\%}$ in the pole mass scheme 
to $\Delta \sigma_{\rm NLO} = {}^{+6.4\%}_{-9.6\%}$ in the \msbar\ scheme.
The above values are taken near the maximum of the curve at $p_T^t = 75~\textrm{GeV}$.

\begin{figure}[t!]
\includegraphics[width=8.5cm]{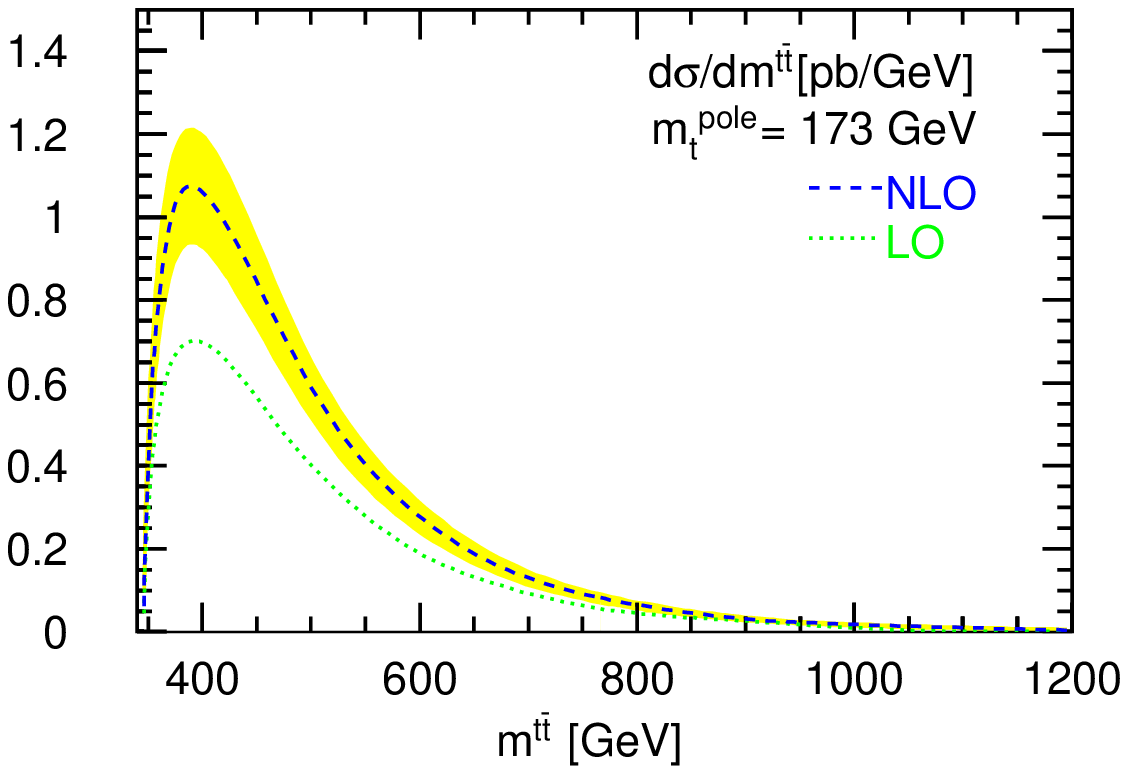}
\includegraphics[width=8.5cm]{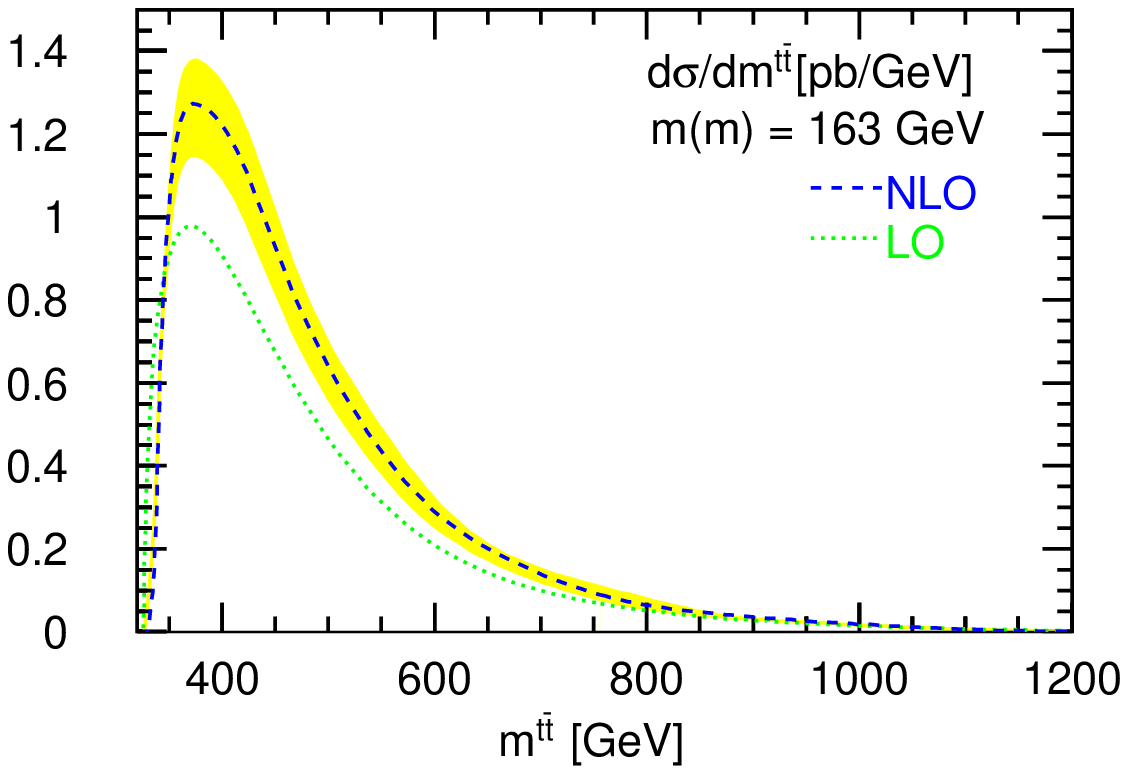}
\caption{\small \label{fig:mtt}
Same as Fig.~\ref{fig:yt} for the differential cross section with
respect to the invariant mass $m^{t{\bar t}}$ of the top quark pair.} 
\end{figure}

Finally, in Fig.~\ref{fig:mtt} we show the invariant mass distributions.
The increase at NLO here is $\sigma_{\rm NLO}/\sigma_{\rm LO} = 1.54$ 
with scale variation $\Delta \sigma_{\rm NLO} = {}^{+13\%}_{-13\%}$ in the pole
mass scheme and $\sigma_{\rm NLO}/\sigma_{\rm LO} = 1.30$ 
with scale variation $\Delta \sigma_{\rm NLO} = {}^{+8.2\%}_{-9.6\%}$ in the \msbar\ scheme.
These values are taken at an invariant mass of $m^{t{\bar t}} = 137~\textrm{GeV}$.

In addition to these improvements, moving from the pole mass to the \msbar\  scheme changes 
the overall shape of the distributions so that the peak positions generally become more pronounced.
This is a consequence of the radiative corrections being shifted to the threshold region as mentioned earlier.
However, the peak positions in both the $p_T^t$ and $m^{t{\bar t}}$ distributions are stable against radiative corrections.
At most they are seen to shift by 1\%, which is unlike the case for $t{\bar t}$-production from $e^+e^-$ collisions 
where the position of the $t{\bar t}$-threshold peak shifts significantly upon 
adding NLO and NNLO perturbative corrections to the total cross section expressed in terms of the pole mass~\cite{Hoang:2000yr}.

Another salient feature not shown in Fig.~\ref{fig:mtt} above occurs 
in the \msbar\  differential cross section with respect to the invariant mass of the $t{\bar t}$ pair.
Very close to the threshold of $t{\bar t}$ production the contribution reponsible for the 
change in the mass renormalization scheme, i.e., the derivative term in eq.~(\ref{eq:crosssecmsb}), becomes large. 
This is due to the presence of a $1/\beta_t$ which diverges as $m^{t{\bar t}} \to m$, cf. eq.~(\ref{eq:pimdefs}).
These large corrections have the effect of causing the invariant mass spectrum
to dip below zero for values of $m^{t{\bar t}} \gsim 2m_t$. 
In the full spectrum, however, this is counterbalanced by the positive contribution resulting 
in a cross section integrated over $m^{t{\bar t}}$ that agrees within less than 1\% with 
the value calculated in {\tt Hathor}.

Obviously, this behavior is an indication of the breakdown of fixed-order perturbation theory.
First of all, bound-state effects in $t{\bar t}$ production at hadron colliders 
arise in the kinematic region $m^{t{\bar t}} \gsim 2m_t$, i.e., 
when the velocity $\beta$ of the top quarks is small, $\beta \ll 1$.
In this region, the conventional perturbative expansion in $\alpha_s$ breaks down, 
owing to singular terms $\sim (\alpha_s/\beta)^n$ in the $n$-loop amplitude,
which require the all-order resummation of the Coulomb corrections~\cite{Hagiwara:2008df,Kiyo:2008bv}.
This resummation for $t{\bar t}$ dynamics close to threshold is carried out in a non-relativistic effective theory 
by means of a Schr\"odinger equation for which the pole mass definition seems to be the natural
choice and which implies a certain power counting, so that all terms of order $m_t\beta^2 \sim m_t\alpha_s^2$
are formally of equal size.

If the contribution for the change in the mass renormalization scheme $\delta m^{\rm sd}$ 
from the pole mass to a so-called short-distance mass $m_t^{\rm sd}$ such as
the \msbar\  mass $m(\mu_r)$ is parametrically larger than $m_t\alpha_s^2$
that is $\delta m^{\rm sd} \equiv m_t^{\rm pole} - m^{\rm sd} \sim m_t^{\rm sd}\alpha_s$, then $\delta m^{\rm sd}$ becomes the dominant term in the kinematic region $m^{t{\bar t}} \gsim 2m_t$.
Such situation is realized for $\delta m^{\rm sd} \sim m_t \alpha_s$, cf. eq.~(\ref{eq:crosssecmsb}),
and excludes the \msbar\  mass from being a useful mass near threshold.
Of course, all these findings on the scheme choice for the mass definition close to the threshold are 
long known from studies for $t{\bar t}$ production in $e^+e^-$ collisions~\cite{Hoang:2000yr}.
Various solutions have been proposed, e.g., the alternative use of a so-called $1S$ mass~\cite{Hoang:1999zc} 
defined through the perturbative contribution to the mass of a hypothetical 
$n = 1$, $^3S_1$ toponium bound state, cf.~\cite{Ahrens:2011px} for an
application to $t{\bar t}$ hadro-production or the use of a ``potential-subtracted'' (PS) mass~\cite{Beneke:1998rk},
recently considered in~\cite{Falgari:2013gwa} in the context of finite-width
effects in unstable-particle production at hadron colliders.
In any case, since the conventional perturbative expansion of the cross
section breaks down for $m^{t{\bar t}} \gsim 2m_t$ we do not display this particular kinematic region in Fig.~\ref{fig:mtt}. 
Moreover, with the currently given experimental resultion for the $m^{t{\bar t}}$-bins, cf.~\cite{Chatrchyan:2012saa},
it will be difficult to access this region at the LHC at all.

For completeness we also provide a table of values for the cross section at LHC
with $\sqrt{S}=8$~TeV at binned values 
of $y^t,p_T^t$ and $m^{t{\bar t}}$ with binning approximately equal to that of \cite{Chatrchyan:2012saa}.
Comparing the data generated using ABM11 as compared to CT10, we see that there 
is an overall shift downward consistent with that observed for the total cross
section, cf. Tabs.~\ref{tab:ttbar-pole} and \ref{tab:ttbar-msbar}.
The improvement of the apparent perturbative convergence and the scale
stability when moving from the pole mass scheme to the \msbar\  scheme is consistent for both PDF sets. 

\begin{table}[th!]
\renewcommand{\arraystretch}{1.3}
\begin{center}
{\small
\hspace*{-5mm}
\begin{tabular}{|c|c|c|c|c|}
\hline
&\multicolumn{2}{|c|}{$m^{\rm pole}_t$ } 
&\multicolumn{2}{|c|}{$m(m)$}
\\     
\hline
$\frac{d\sigma}{dy^t}$ &
LO & NLO &
LO & NLO
\\
\hline
$y^t = 0.2 $ & 48.70 & 73.43 & 64.46 & 84.83
\\
$y^t = 0.6 $ & 44.12 & 66.34 & 58.57 & 76.74
\\
$y^t = 1.0 $ & 35.90 & 53.70 & 48.00 & 62.29
\\
$y^t = 1.4 $ & 25.77 & 38.19 & 34.87 & 44.51
\\
$y^t = 2.0 $ & 11.37 & 16.39 & 15.93 & 19.34
\\
\hline
\end{tabular}
}
\caption{\small 
Values for the $y^t$ differential cross section for top-quark pair-production at LO and NLO for various $y^t$ using the PDF set CT10~\cite{Gao:2013xoa} with $\sqrt{S} = 8$TeV.
All rates are in pb. 
}
\label{tab:yCT10}
\end{center}
\end{table}

\begin{table}[th!]
\renewcommand{\arraystretch}{1.3}
\begin{center}
{\small
\hspace*{-5mm}
\begin{tabular}{|c|c|c|c|c|}
\hline
&\multicolumn{2}{|c|}{$m^{\rm pole}_t$ } 
&\multicolumn{2}{|c|}{$m(m)$}
\\     
\hline
$\frac{d\sigma}{dp_T^t}$ &
LO & NLO &
LO & NLO
\\
\hline
$y^t = 0.2 $ & 44.39 & 65.82 & 59.51 & 76.33
\\
$y^t = 0.6 $ & 39.55 & 58.57 & 53.18 & 68.00
\\
$y^t = 1.0 $ & 31.07 & 45.89 & 42.06 & 53.44
\\
$y^t = 1.4 $ & 21.04 & 30.91 & 28.83 & 36.18
\\
$y^t = 2.0 $ & 8.018 & 11.55 & 11.40 & 13.72
\\
\hline
\end{tabular}
}
\caption{\small 
The same as table \ref{tab:yCT10} but using the PDF set ABM11~\cite{Alekhin:2012ig} .
}
\label{tab:yABM11}
\end{center}
\end{table}

\begin{table}[th!]
\renewcommand{\arraystretch}{1.3}
\begin{center}
{\small
\hspace*{-5mm}
\begin{tabular}{|c|c|c|c|c|}
\hline
&\multicolumn{2}{|c|}{$m^{\rm pole}_t$ } 
&\multicolumn{2}{|c|}{$m(m)$}
\\     
\hline
$\frac{d\sigma}{dp_T^t}$ &
LO & NLO &
LO & NLO
\\
\hline
$p_T^t = 30 {\tt GeV}$ & 0.5513 & 0.8681 & 0.8214 & 1.058
\\
$p_T^t = 90 {\tt GeV}$ & 0.9364 & 1.399 & 1.308 & 1.637
\\
$p_T^t = 130 {\tt GeV}$ & 0.7130 & 1.045 & 0.9419 & 1.196
\\
$p_T^t = 170 {\tt GeV}$ & 0.4422 & 0.6288 & 0.5455 & 0.7057
\\
$p_T^t = 230 {\tt GeV}$ & 0.1777 & 0.2496 & 0.2070 & 0.2675
\\
$p_T^t = 290 {\tt GeV}$ & 0.06806 & 0.09941 & 0.08152 & 0.1035
\\
$p_T^t = 360 {\tt GeV}$ & 0.02533 & 0.03105 & 0.02756 & 0.03537
\\
\hline
\end{tabular}
}
\caption{\small 
Values for the $p_T^t$ differential cross section for top-quark pair-production at LO and NLO for various $p_T^t$ using the PDF set CT10~\cite{Gao:2013xoa}.
All rates are in pb/GeV. 
}
\label{tab:ptCT10}
\end{center}
\end{table}

\begin{table}[th!]
\renewcommand{\arraystretch}{1.3}
\begin{center}
{\small
\hspace*{-5mm}
\begin{tabular}{|c|c|c|c|c|}
\hline
&\multicolumn{2}{|c|}{$m^{\rm pole}_t$ } 
&\multicolumn{2}{|c|}{$m(m)$}
\\     
\hline
$\frac{d\sigma}{dp_T^t}$ &
LO & NLO &
LO & NLO
\\
\hline
$p_T^t = 30 {\tt GeV}$ & 0.4874 & 0.7568 & 0.7467 & 0.9220
\\
$p_T^t = 90 {\tt GeV}$ & 0.8141 & 1.206 & 1.148 & 1.429
\\
$p_T^t = 130 {\tt GeV}$ & 0.6076 & 0.8862 & 0.8053 & 1.006
\\
$p_T^t = 170 {\tt GeV}$ & 0.3658 & 0.5262 & 0.4429 & 0.5843
\\
$p_T^t = 230 {\tt GeV}$ & 0.1425 & 0.1954 & 0.1750 & 0.2175
\\
$p_T^t = 290 {\tt GeV}$ & 0.05567 & 0.06975 & 0.06227 & 0.07316
\\
$p_T^t = 360 {\tt GeV}$ & 0.02008 & 0.02415 & 0.01266 & 0.01818
\\
\hline
\end{tabular}
}
\caption{\small 
The same as table \ref{tab:ptCT10} but using the PDF set ABM11~\cite{Alekhin:2012ig} .
}
\label{tab:ptABM11}
\end{center}
\end{table}

\begin{table}[th!]
\renewcommand{\arraystretch}{1.3}
\begin{center}
{\small
\hspace*{-5mm}
\begin{tabular}{|c|c|c|c|c|}
\hline
&\multicolumn{2}{|c|}{$m^{\rm pole}_t$ } 
&\multicolumn{2}{|c|}{$m(m)$}
\\     
\hline
$\frac{d\sigma}{dm^{t{\bar t}}}$ &
LO & NLO &
LO & NLO
\\
\hline
$m^{t{\bar t}} = 350 {\tt GeV}$ & 0.2985 & 0.4278 & 0.9046 & 1.0295
\\
$m^{t{\bar t}}  = 450 {\tt GeV}$ & 0.5648 & 0.8441 & 0.6755 & 0.9270
\\
$m^{t{\bar t}}  = 500 {\tt GeV}$ & 0.4022 & 0.5914 & 0.4656 & 0.6403
\\
$m^{t{\bar t}}  = 600 {\tt GeV}$ & 0.1898 & 0.2782 & 0.2102 & 0.2917
\\
$m^{t{\bar t}}  = 700 {\tt GeV}$ & 0.09342 & 0.1301 & 0.09977 & 0.1404
\\
$m^{t{\bar t}}  = 950 {\tt GeV}$ & 0.01796 & 0.02343 & 0.02067 & 0.02740
\\
\hline
\end{tabular}
}
\caption{\small 
Values for the $m^{t{\bar t}}$ differential cross section for top-quark pair-production at LO and NLO for various $m^{t{\bar t}}$ using the PDF set CT10~\cite{Gao:2013xoa}.
All rates are in pb/GeV. 
}
\label{tab:mttCT10}
\end{center}
\end{table}
\begin{table}[th!]
\renewcommand{\arraystretch}{1.3}
\begin{center}
{\small
\hspace*{-5mm}
\begin{tabular}{|c|c|c|c|c|}
\hline
&\multicolumn{2}{|c|}{$m^{\rm pole}_t$ } 
&\multicolumn{2}{|c|}{$m(m)$}
\\     
\hline
$\frac{d\sigma}{dm^{t{\bar t}}}$ &
LO & NLO &
LO & NLO
\\
\hline
$m^{t{\bar t}} = 350 {\tt GeV}$ & 0.3036 & 0.4546 & 0.8420 & 0.9508
\\
$m^{t{\bar t}}  = 450 {\tt GeV}$ & 0.4967 & 0.7381 & 0.5914 & 0.8103
\\
$m^{t{\bar t}}  = 500 {\tt GeV}$ & 0.3481 & 0.5118 & 0.3964 & 0.54488
\\
$m^{t{\bar t}}  = 600 {\tt GeV}$ & 0.1554 & 0.2212 & 0.1704 & 0.2357
\\
$m^{t{\bar t}}  = 700 {\tt GeV}$ & 0.0729 & 0.09674 & 0.07706 & 0.1061
\\
$m^{t{\bar t}}  = 950 {\tt GeV}$ & 0.01326 & 0.01839 & 0.01407 & 0.01611
\\
\hline
\end{tabular}
}
\caption{\small 
The same as table \ref{tab:mttCT10} but using the PDF set ABM11~\cite{Alekhin:2012ig} .
}
\label{tab:mttABM11}
\end{center}
\end{table}

\bigskip

In summary, we have shown how treating the differential cross sections for $t{\bar t}$ production 
in the \msbar\  scheme for the top-quark mass has benefits as compared to the pole mass scheme.
The perturbative series shows the same improvement in convergence and scale dependence as has been observed for the total cross section.
As a consequence the NLO contributions with a \msbar\  mass are expected to provide already very precise cross section predictions.
An extension to NNLO accuracy would provide results with a still smaller theoretical uncertainty from the scale variation. 
Yet, the predictions at the nominal scale, i.e., $\mu_r=m(m)$, are expected to remain largely unchanged.

As future prospects we note that the refinement of the present phenomenological analysis 
to NNLO accuracy is certainly feasible once the complete NNLO QCD corrections
for differential $t{\bar t}$ production are available.
As a first step in this direction, one may consider approximate NNLO
corrections based, e.g., on the dominant threshold logarithms.
Other obvious improvements are extension to double-differential distributions
and other exclusive observables, even including top-quark decay.

\subsection*{Acknowledgments}
This work is partially supported by the Deutsche Forschungsgemeinschaft in Sonderforschungs\-be\-reich/Transregio~9
and by the European Commission through contract PITN-GA-2010-264564 ({\it LHCPhenoNet}).

{\footnotesize


}

\end{document}